\documentclass[fleqn,usenatbib]{mnras}
\usepackage{newtxtext,newtxmath}
\usepackage[T1]{fontenc}
\DeclareRobustCommand{\VAN}[3]{#2}
\let\VANthebibliography\thebibliography
\def\thebibliography{\DeclareRobustCommand{\VAN}[3]{##3}\VANthebibliography}
\usepackage{graphics}
\usepackage{tabu}
\usepackage{natbib}
\usepackage{color}
\usepackage{rotating}
\usepackage{multirow}
\usepackage{footnote}
\usepackage{datetime}
\usepackage{amsmath}
\usepackage{mathrsfs}
\usepackage[normalem]{ulem}
\usepackage{subfigure}

\usepackage{lipsum} 

\title[Compact Galaxy SMBH]{Black Hole Feeding and Feedback in a Compact Galaxy}

\author[Y. Di et al.]{
Yihuan Di$^{1,2}$, 
Yuan Li$^{3}$\thanks{E-mail: yuan.li@unt.edu}, 
Feng Yuan$^{1,2}$\thanks{E-mail:fyuan@shao.ac.cn}, Fangzheng Shi$^1$, and
Mirielle Caradonna$^{3}$\\ 
$^{1}$Shanghai Astronomical Observatory 
Chinese Academy of Sciences, 80 Nandan Road, Shanghai 200030, People’s Republic of China\\
$^{2}$University of Chinese Academy of Sciences, 19A Yuquan Road, Beijing 100049, People’s Republic of China\\
$^{3}$Department of Physics, University of North Texas, Denton, TX 76203, USA\\}

\date{Accepted XXX. Received YYY; in original form ZZZ}

\pubyear{2022}

\begin{document}
\label{firstpage}
\pagerange{\pageref{firstpage}--\pageref{lastpage}}
\maketitle

\begin{abstract}  
We perform high-resolution hydrodynamical simulations using the framework of {\it MACER} to investigate supermassive black hole (SMBH) feeding and feedback in a massive compact galaxy, which has a small effective radius but a large stellar mass , with a simulation duration of 10 Gyr. We compare the results with a reference galaxy with a similar stellar mass but a less concentrated stellar density distribution, as typically found in local elliptical galaxies. 
We find that about 10\% of the time, the compact galaxy develops multi-phase gas within a few kpc, 
but the accretion flow through the inner boundary below the Bondi radius is always a single phase. The inflow rate in the compact galaxy is several times larger than in the reference galaxy, mainly due to the higher gas density caused by the more compact stellar distribution. Such a higher inflow rate results in stronger SMBH feeding and feedback and a larger fountain-like inflow-outflow structure. Compared to the reference galaxy, the star formation rate in the compact galaxy is roughly two orders of magnitude higher but is still low enough to be considered quiescent. Over the whole evolution period, the black hole mass grows by $\sim$50\% in the compact galaxy, much larger than the value of $\sim$ 3\% in the reference galaxy.    

\end{abstract}



\section{Introduction}
\label{sec:intro} 
\setcounter{footnote}{0} 

Observations suggest that the progenitors of today's massive elliptical galaxies were compact quiescent galaxies at $z\sim 2$, often termed ``red nuggets'' \citep[e.g.,][]{2009ApJ...700..221K, 2009ApJ...695..101D}. Numerical simulations show that they are formed through a compaction-quenching process \citep{2013ApJ...776...63F, 2015MNRAS.450.2327Z, 2015Sci...348..314T}. At lower $z$, the evolution of massive early-type galaxies is dominated by minor mergers which mainly add stars to the outskirts of the galaxies to grow their sizes  \citep{2010ApJ...709.1018V,Oser2010,2009ApJ...699L.178N}.  Recent observations have revealed a small population of local compact galaxies with small effective radii but large stellar masses \citep{2017MNRAS.468.4216Y}. These compact galaxies are suggested to be relics of the $z\sim 2$ red nuggets that have evolved passively in isolation and thus did not experience as many late-time mergers to grow their sizes \citep{2014ApJ...780L..20T}. Like the more typical massive elliptical galaxies, these red nugget relics also show little to no ongoing star formation and are surrounded by hot X-ray halos \citep{2018ApJ...854..143B, 2018MNRAS.477.3886W}.

It is generally accepted that the quiescent state of massive elliptical galaxies is maintained by the feedback from the supermassive black holes (SMBHs) in the centers of galaxies \citep{2007ARA&A..45..117M, 2012ARA&A..50..455F}. The roles of SMBHs in galaxy evolution have been studied extensively in large-scale cosmological simulations \citep[e.g.,][]{2015MNRAS.446..521S, 2018MNRAS.479.4056W}. 
Idealized galaxy-scale simulations have explored more details of the feeding and feedback of SMBHs thanks to the higher spatial resolution they are able to achieve \citep[e.g.,][]{2010ApJ...717..708C,Yuan2018,Wang2019, 2020ApJ...905...50P}. 
However, almost all previous theoretical studies have been focused on typical massive elliptical galaxies. How SMBH feeding and feedback operate in compact galaxies and how they differ from typical massive galaxies are not well-understood from a theoretical perspective. 

In this work, we explore SMBH feeding and feedback in an idealized, isolated compact galaxy. We also perform a simulation of a typical low z massive elliptical galaxy for comparison. The two-dimensional hydrodynamical simulations are performed using the {\it MACER} framework \citep{Yuan2018}, resolving from galaxy scales all the way down below the Bondi radius. In section \ref{sec:metho}, we present a brief overview to {\it MACER} and the setup of our simulations. Our main results are discussed in Section~\ref{sec:results}, including the accretion flow, SMBH feedback, and star formation in the simulations. We conclude our work in Section~\ref{sec:conclusions}.

\begin{samepage}
\section{Methodology} \label{sec:metho}
\subsection{The {\it MACER} framework: overview}
\end{samepage}
The simulations in this paper are performed using MACER (Massive AGN Controlled Ellipticals Resolved), a high-resolution hydrodynamical numerical simulation framework built on the ZEUS code \citep{1992ApJS...80..753S} developed over the past decades to study the evolution of elliptical galaxies, emphasizing the role of AGN feedback  \citep[e.g.,][]{2001ApJ...551..131C,2010ApJ...717..708C,Yuan2018}. 
The detailed descriptions of the framework are presented in \citet{Yuan2018}. Here we briefly introduce the key features of {\it MACER} (this section) and the AGN physics adopted in it (section \ref{sec:Phy}).

{\it MACER} uses spherical coordinate ($r,\theta,\phi$).  Axial symmetry is assumed in the $\phi$-direction and we focus only on the ($r,\theta$) plane; thus our simulation is 2.5 dimensional.  The inner boundary of the simulation is adopted to be small enough to resolve the  Bondi radius, $r_{\rm in}\la r_{\rm Bondi}$, so that we can reliably calculate the mass flux there.  We can then combine the calculated mass flux with the theory of black hole accretion to obtain the value of the mass accretion rate at the black hole horizon and properties of radiation and wind. The value of mass accretion rate is crucial to the AGN feedback study since it determines the power of the AGN. Due to limited resolution, cosmological simulations cannot resolve the Bondi radius thus have to adopt various approximated approaches to estimate the accretion rate, whose uncertainty is found to be as large as $\sim 300$ \citep{2017MNRAS.467.3475N}. 

The second characteristic of our model is that we adopt the state-of-the-art physics of black hole accretion. Especially, the results of wind and radiation in the hot accretion mode developed in the recent decade are incorporated, as we will briefly introduce in section \ref{sec:Phy}. 
We inject radiation and wind at the inner boundary $r_{\rm in}$ and calculate their energy and momentum interactions with the gas in the galaxy. For AGN wind, we treat it as a source term by adding the energy, momentum and mass of the wind into the innermost two grids of the simulation domain. For AGN radiation, we consider the heating/cooling and radiation pressure to the gas. The radiative heating and cooling processes include bremsstrahlung, Compton heating and cooling, photoionization heating, line and recombination continuum cooling. 
We then self-consistently calculate the interaction between AGN outputs and the interstellar medium. In contrast, in cosmological models, phenomenological parameters are usually adopted to couple the AGN outputs with the gas in an assumed region surrounding the AGN. 

\subsection{Physics of AGN feedback adopted in {\it MACER}}
\label{sec:Phy}

Motivated by the observations and theoretical models of black hole X-ray binaries, black hole accretion comes in hot and cold modes, bounded by $2\%L_{\rm Edd}$ \citep{Yuan2014}. The existence of two accretion modes and the value of this  ``boundary luminosity''  between them should also apply to the supermassive black holes in the centers of galaxies, because the physics of black hole accretion is largely independent of black hole mass. 
When the accretion is in the cold mode, the inflowing gas first freely falls until a disk is formed at the circularization radius determined by the specific angular momentum of the falling gas. The black hole accretion rate $\dot{M}_{\rm BH}$ is calculated from the accretion rate at the inner boundary by solving a set of differential equations, taking into account the mass evolution of the small disk and the mass lost in the wind. The bolometric luminosity is calculated by $L_\mathrm{bol} = \epsilon_\mathrm{cold} \dot{M}_\mathrm{BH}c^2$, with $\epsilon_\mathrm{cold}=0.1$ is the radiative efficiency. The mass flux and velocity of wind launched in the cold mode as a function of $L_\mathrm{bol}$ are taken from observational results of \citet{Gofford2015}. The mass flux of wind is assumed to be $\propto {\rm cos}^2(\theta)$, i.e., the wind has a bipolar structure \citep{Yuan2018}.

When the accretion is in the hot mode, the accretion flow consists of an outer truncated thin disk and an inner hot accretion flow, with the value of the truncation radius determined by the mass accretion rate at $r_{\rm in}$ and the black hole mass \citep{Yuan2014}. 
The dynamics and radiation of the hot accretion flow have been intensively studied in the past decades \citep[see review by][]{Yuan2014}. Different from the thin disk, the radiative efficiency of a hot accretion flow is a function of accretion rate \citep{Xie2012}.

 Numerical simulations have shown that strong wind must be produced throughout the hot accretion flows \citep{2012ApJ...761..129Y,2012ApJ...761..130Y,2012MNRAS.426.3241N,Yuan2015,2016ApJ...818...83B,2016ApJ...823...90B,Yang2021}. The prediction  has been confirmed by more and more observations in recent years \citep{wang2013,Cheung2016,Park2019,2019MNRAS.483.5614M,Shi2021,2022ApJ...926..209S}. Since the observational constraints on the properties of hot wind are still weak, in {\it MACER}, we take the properties of the wind, including its mass flux, velocity, and spatial distribution, from the $\phi-$averaged results of three-dimensional general relativity MHD numerical simulation of black hole accretion flows of \citet{Yuan2015}. For simplicity, we assume that the accretion is in the SANE  (standard and normal evolution) mode and the black hole spin $a=0$. 

\subsection{Physics of stellar evolution adopted in {\it MACER}}

In our simulation, the gas is mainly supplied by stellar evolution. The total mass-loss rate of a stellar population is
\begin{equation}
\dot{M}(t)=\dot{M}_\star(t) + \dot{M}_{SN}(t).
\end{equation}
Here $\dot{M}_{\star}$ is the mass-loss rate, which is approximated as ``simple stellar population'', namely all stars are formed at the same time. It is determined by \citep{2005MNRAS.362..799M},
\begin{equation}
\dot{M}_{\star}=10^{-12}A M_{\star}t^{-1.3}_{12}\ \mathrm{M_\odot yr^{-1}},
\end{equation}
where $M_{\star}$ is the stellar mass of the galaxy at an age of 12 Gyr, and $t_{12}$ is the age in units of 12 Gyr. The coefficient $A$ is set to 3.3, which is Kroupa initial mass function. The recycle rate of gas from SN Ia is $\dot{M}_{SN} = 1.4 \mathrm{M_{\odot}}R_{SN}(t)$, where $R_{SN}$ is the evolution of the explosion rate with time. 

The star formation rate is calculated as \citep{Yuan2018}:
\begin{equation}
   \dot{\rho}_{\rm SF}=\eta_{\rm SF}\frac{\rho}{\tau_{\rm SF}}, 
\end{equation}
with $\eta_{\rm SF}$ being the efficiency. The calculation of the timescale $\tau_{\rm SF}$ can be found in \citet{Yuan2018}.  But different from \citet{Yuan2018}, here we adopt thresholds for the density and temperature, which is widely adopted in numerical simulations due to insufficient resolutions. We assume that gas can be converted into stars only when its density is higher than $1~{\rm cm}^{-3}$ and the temperature is lower than $4\times 10^4$K.

The star formation removes mass, momentum, and energy of gas from the grid, while SNe II injects mass and energy. The newly formed star is assumed to have a Salpeter IMF, the mass return from SN II progenitors is 20\% of the newly formed stellar mass.

\subsection{Galaxy initial conditions}\label{sec:initial}

\begin{figure}
\centering
\includegraphics[width=8.2cm]{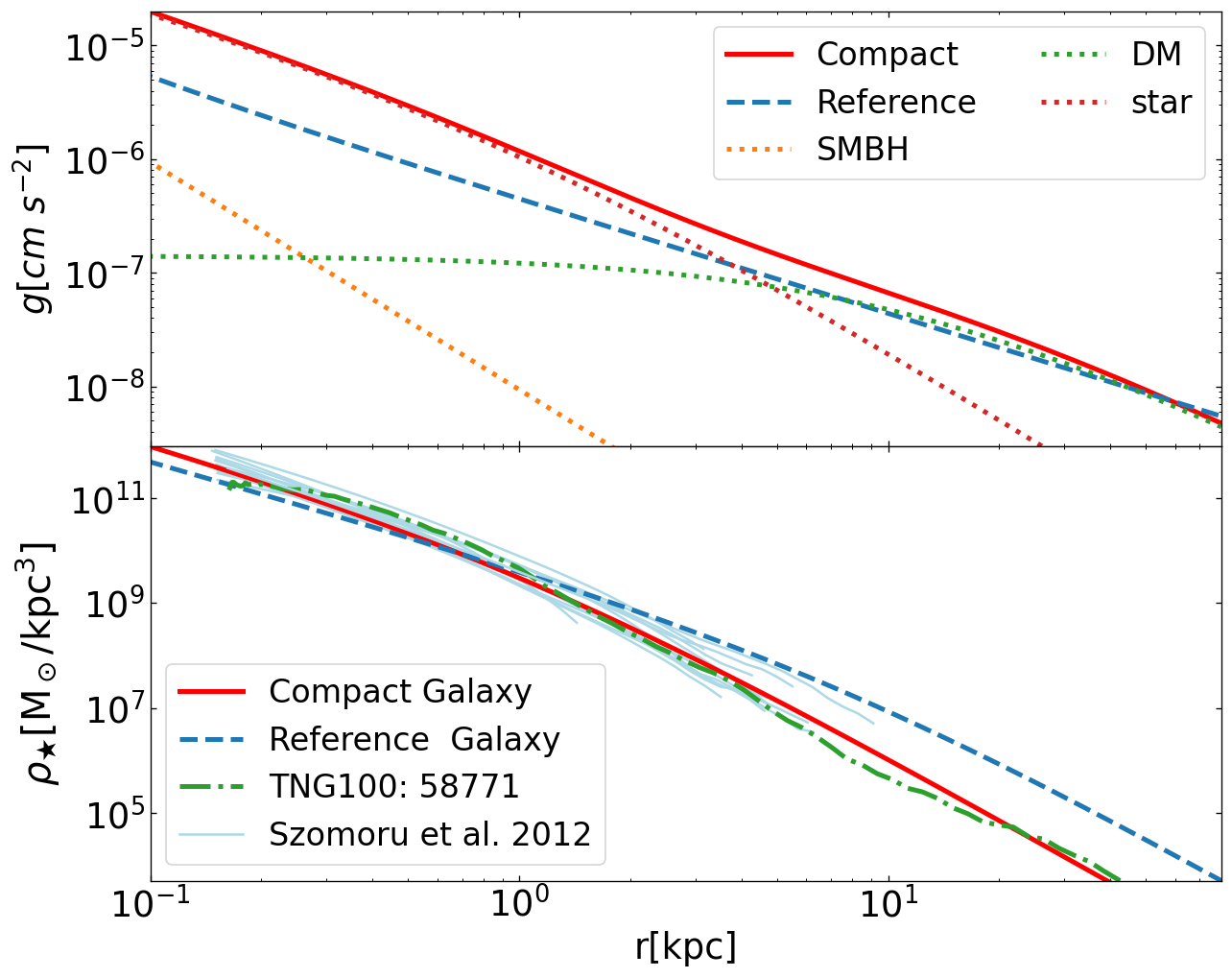}
\caption{ {\it Top panel:} The gravitational potential of the compact galaxy and the reference galaxy. {\it Bottom panel:} The stellar profile of the compact (red solid) and reference (blue dashed) galaxies used in this work. The light blue lines show the profiles of a sample of compact massive galaxies at $z \sim 2$, taken from \citet{Szomoru2012}. The green line shows the original data from IllustrisTNG.}
\label{fig:stellar-profile}
\end{figure}

The initial condition of our compact galaxy is based on a compact galaxy (ID: 58771) extracted from the IllustrisTNG simulation \citep{2018MNRAS.473.4077P} at $z=$ 2. The galaxy is selected using the criteria from \cite{Dokkum2015}:
\begin{equation}
    \mathrm{log}(R_\mathrm{e}/\mathrm{kpc}) \leq \mathrm{log}(M_*/\mathrm{M_\odot}) -10.7,
\end{equation}
with $\log(M_{*}/\mathrm{M_\odot}) \ge 10.6 $, where $M_{*}$ is the stellar mass and $R_\mathrm{eff}$ is effective radius. We fit the dark matter halo using an NFW profile \citep{Navarro1996} with a scale radius $r_s=8.268$ kpc and a dark matter mass of $M_{\rm vir}=2.98 \times 10^{12} \mathrm{M_\odot}$.
The stellar density profile is fitted with a spherically symmetric Jaffe model \citep{Dehnen1993}:
\begin{equation}
\rho_*=\frac{M_*}{4\pi}\times \frac{a}{r_\mathrm{eff}^2 \times \left(r_\mathrm{eff}+a\right)^2} \,,
\label{Jaffe}
\end{equation}
with an effective radius $r_\mathrm{eff}=1.03 \mathrm{kpc}$ and a stellar mass $M_*=1.52\times 10^{11} \mathrm{M_\odot}$. 
The values of these two parameters are obtained by fitting the stellar distribution in the IllustrisTNG simulation data.  We use the $\beta$ model for the initial gas density profile:
\begin{equation}
\rho_\mathrm{gas}=\rho_{0} \times \left(1+\left(\frac{r}{r_\mathrm{c}}\right)^2\right)^{-\frac{3}{2} \beta}    \,, 
\end{equation}
with $ \beta = 0.8$ and $r_\mathrm{c} = 8.1614$ kpc. Again, the values of these two parameters are obtained by fitting the gas distribution in the IllustrisTNG simulation data. The temperature profile is obtained by hydrostatic equilibrium calculation. We find that when our simulations reach the quasi-steady state, the gas within $\sim 10$ kpc is dominated by the mass loss from the stellar evolution; while beyond this radius, the gas is dominated by the ``initial'' gas. The central black hole has an initial mass $M_\mathrm{BH} = 6.723\times 10^8 \mathrm{M_\odot}$.

For reference, we also simulate a typical local massive elliptical galaxy. Its parameters are described in \cite{Yuan2018}, except that a same black hole mass as the compact galaxy is adopted for comparison purpose. The stellar profile of this reference galaxy is described by a Jaffe model, with $M_*=3 \times 10^{11} M_{\odot}$ and $r_{\rm eff}=6.9$ kpc (refer to Eq. \ref{Jaffe}). The stellar profiles of the compact and reference galaxies are shown in Fig. \ref{fig:stellar-profile}, together with the initial condition taken from IllustrisTNG. The density profile of the sum of stellar matter and dark matter is given by 
\begin{equation}
    \rho_\mathrm{T} = \frac{\sigma^2_0}{2 \pi Gr^2}.
\end{equation}
Here $\sigma_0$  is the projected central velocity dispersion ($\sigma_0=260 {\rm km/s}$). The initial gas is neglected as we find they are quickly ejected out and replaced by the gas that originated from stellar evolution. 

The assumed specific angular momentum of the stars providing gas in the simulation can be calculated from this equation \citep{2011ApJ...737...26N}:
\begin{equation}
  \frac{1}{v_{\phi}(R)}=\frac{d}{\sigma R}+\frac{1}{f\sigma}+\frac{R}{j},  
\end{equation}
where $R$ is the distance to the $z$-axis, $\sigma$ is the central one-dimensional line-of-sight velocity dispersion for the galaxy model, and $d,f$, and $j$ are adjustable parameters controlling the angular momentum profile at various radii. 
In our simulation, the value of $v_\phi$ at the innermost radius is only in the tens of ${\rm km/s}$.
Both the compact massive galaxy and the reference galaxy are assumed to be isolated and slowly rotating.

\subsection{Model setting}
The inner and outer boundaries of the simulation domain are set to be 0.8pc and 160 kpc. We have calculated the mass flux-weighted Bondi radius and found that this value of the inner boundary is $2\sim 10$ times smaller than the Bondi radius. Our fiducial simulations have $120\times 30$ grids in the $r-\theta$ plane. In the $\theta$ direction, the mesh is divided homogeneously; in the radial direction, we use a logarithmic mesh. With such grids, the finest resolution is achieved in the innermost region, which is $\sim$0.17 pc.

\section{Results and Discussions}\label{sec:results}
\subsection{Gas Motion from Galaxy Scale to Bondi Radius}\label{sec:accretion}

\begin{figure}  
\centering
\includegraphics[width=8.2cm,trim=0.3cm 0.1cm 0.2cm 0cm, clip=true]{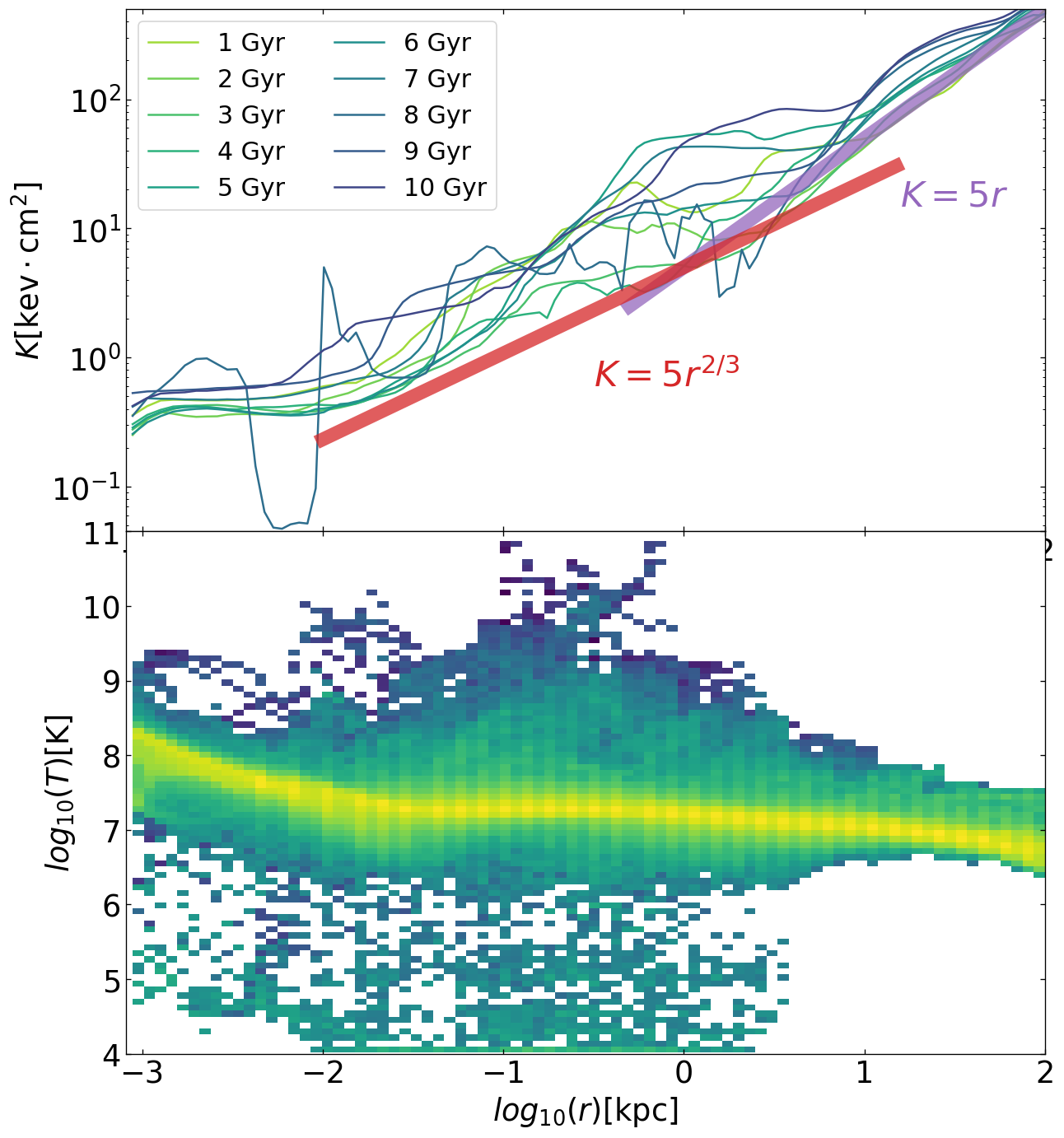}
\caption{{\it  Top panel:} Evolution of the gas entropy profiles in the simulated compact galaxy over 10 Gyr, sampled every Gyr (from yellow to blue). Also included are the entropy profiles from structure formation (purple) and the precipitation threshold (red). The large dip in the blue line around $log(r)=-2$ is due to the presence of a dense cold clump. {\it Bottom panel:} Gas temperature distribution stacked over the entire duration of the simulation. Condensation occurs from $\sim 10$ pc to $\sim 2$ kpc. But the cool gas is dissolved before reaching the inner boundary, and the accretion flow is hot there. }
\label{fig:Tr}
\end{figure}

The interplay between radiative cooling and AGN feedback regulates the simulated galaxy to a quasi-steady state. The top panel of Figure~\ref{fig:Tr} shows the evolution of the hot gas entropy profiles. Entropy $K$ here is defined as $K\equiv kTn_e^{-2/3}$. On large scales ($r>2$ kpc), the entropy profile roughly follows $r$, which is the baseline entropy profile from gravitational structure formation \citep{Voit2005}. From $\sim 10$ pc to $\sim 2$ kpc, the entropy profile scales as $r^{2/3}$, following the precipitation limit \citep{Voit2015}. Such an entropy limit is equivalent to the $t_{cool}/t_{ff} \sim 10$ criteria \citep{Sharma2012, 2014ApJ...789..153L,Ji2018}. At the innermost region ($1-10$ pc), the entropy is roughly flat. Previous simulations with lower resolution typically show a much larger flat entropy core of kpc scales \citep[e.g.,][]{Voit2017}. Our smaller entropy core ($<100$ pc) is more consistent with the observed entropy profiles of elliptical galaxies \citep{Voit2015}.

The bottom panel of Figure~\ref{fig:Tr} shows the stacked gas temperature as a function of $r$. Gas condensation indeed develops from $\sim 10$ pc to $\sim 2$ kpc, where the entropy of the hot gas follows the precipitation limit. Previous simulations suggest that the accretion of cold gas dominates at the Bondi radius when multiphase gas is present \citep{Gaspari2013}. However, when even smaller scales are resolved, we find that cool gas is dissolved before reaching the smallest radius, and the accretion flow is hot. This is consistent with the recent findings in \citet{Guo2022} for simulated three-dimensional accretion flows in M87. We note that although the galaxy develops multi-phase gas sometimes, it is a single-phase galaxy about 90\% of the time in the simulation, similar to the simulated reference galaxy. 

\begin{figure*}  
\centering
\includegraphics[width=15cm]{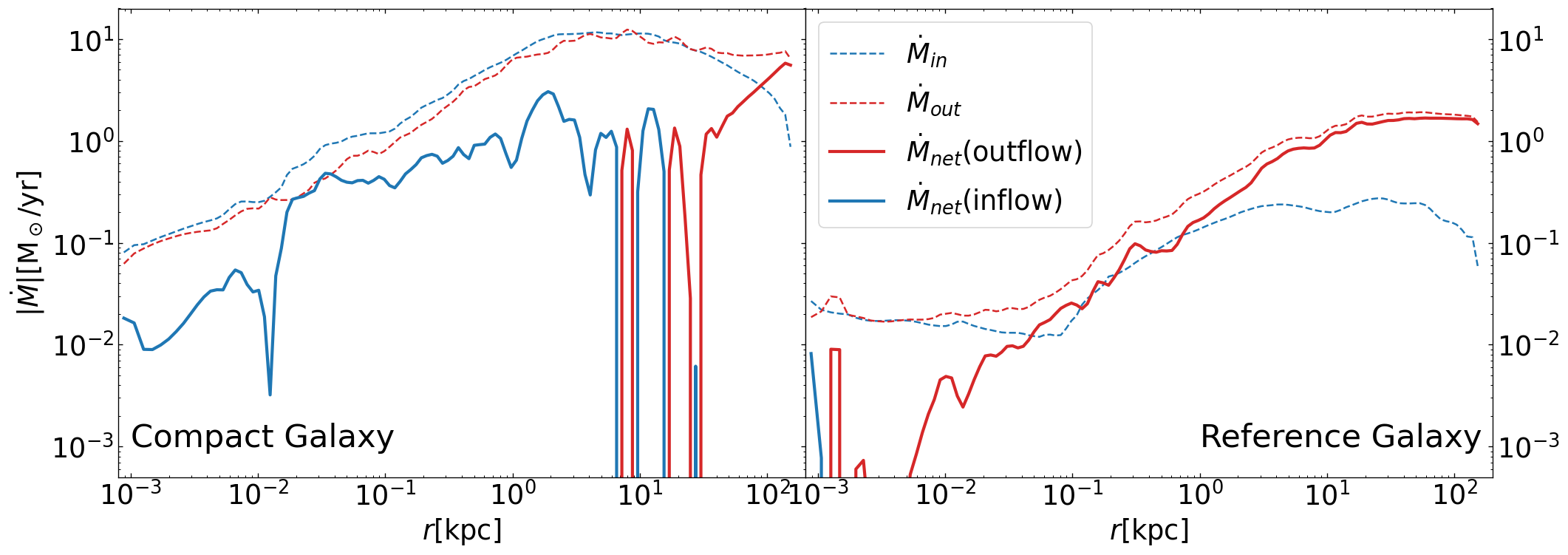}
\caption{Radial profiles of time- and angle-averaged inflow, outflow, and net rates for the compact (left) and reference (right) galaxies over the whole evolution period. The net rates shown are the absolute values. 
}
\label{fig:inflow}
\end{figure*}

 We define the time and angle-averaged inflow, outflow, and net rates as \citet{Ressler2018}:
\begin{align}
\dot{M}_{\mathrm{in}} & \equiv-\left\langle 4 \pi \rho \min \left(v_{r}, 0\right) r^{2}\right\rangle\notag \\
    \dot{M}_{\mathrm{out}} & \equiv\left\langle 4 \pi \rho \max \left(v_{\mathrm{r}}, 0\right) r^{2}\right\rangle\notag \\
    \dot{M}_{\mathrm{net}} & \equiv \dot{M}_{\mathrm{in}}-\dot{M}_{\mathrm{out}}
\end{align}
The rates are averaged over 10 Gyr. Figure~\ref{fig:inflow} shows the radial profiles of the three rates for compact and reference galaxies.
We find that for both galaxies, both the inflow and outflow rates decrease with decreasing radius. Such an inward decrease is similar to that of the black hole hot accretion flow \citep{2012ApJ...761..129Y}. We speculate that, as in the case of the black hole accretion flow, it may be due to the presence of convective motion and/or outflow in the galaxy \citep{2012ApJ...761..130Y}. One important difference is that here the net rate is not a constant of radius. This is because we have gas sources from stellar evolution, which is not considered in the case of black hole accretion flow.  We note that our result is similar to what is found in \citet{Ressler2018} for the Milky Way SMBH fed by stellar winds. 

The location of the ``stagnation region'', where the inflow and outflow have comparable strengths and thus the net flow rapidly switches between the two, is very different in the two simulated galaxies. In the compact galaxy, it is located around $\sim 10$ kpc; while in the reference galaxy, it is three orders of magnitude closer to the SMBH, at only a few pc. Within a few kpc, the compact galaxy is entirely dominated by inflow; whereas in the reference galaxy, inflow only takes over close to the inner boundary. As shown by the top panel of Figure \ref{fig:stellar-profile}, the gravitational force in the compact galaxy is typically stronger than that in the reference galaxy by a factor of a few. This is why the inflow region in the former is much larger than in the latter.  

At galaxy scales, ranging from the inner boundary to $\sim 10$ kpc, the magnitude of the inflow and outflow rates in the compact galaxy is several times higher than those in the reference galaxy. This is mainly because the gas density in the central region of the compact galaxy is a few times higher than the reference galaxy. The stellar wind is the main source of gas in the galaxy. The higher density in the compact galaxy is because its stellar distribution is more compact. 

This higher inflow rate at the inner boundary implies stronger feeding, and subsequently feedback, of the SMBH, which we discuss in more detail in Section~\ref{sec:feedback}.

When comparing the outflow rate and the stellar mass loss from evolved stars, we find that in the reference galaxy, the two are comparable. This suggests that within a few kpc, the outflow is mainly composed of stellar wind. In the compact galaxy, stellar wind makes up a small fraction of the inflow and outflow rates. This suggests that the stellar wind gas is difficult to escape out of the galaxy due to the strong gravity of the galaxy; rather, the inflow and outflow motions are likely dominated by the strong galaxy-scale circulation.

\subsection{SMBH Feedback}\label{sec:feedback}

\begin{figure*}  
\centering
\subfigure[Averaged over 10 Gyr period]{
\label{fig:avr}
\includegraphics[width=7.4cm]{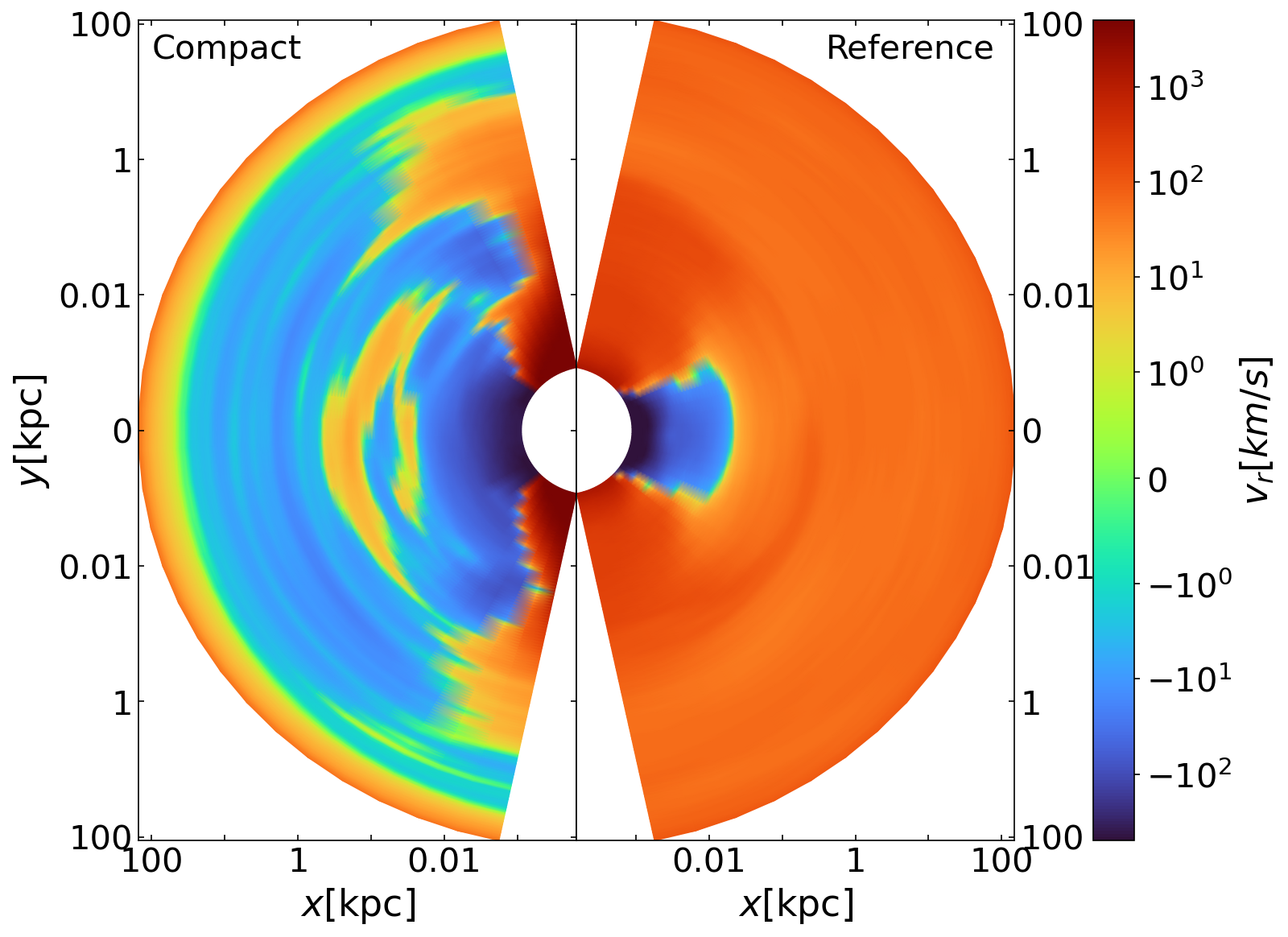}}
\subfigure[Snapshot at 9.75 Gyr in 10 kpc]{
\label{fig:snap}
\includegraphics[width=7.4cm]{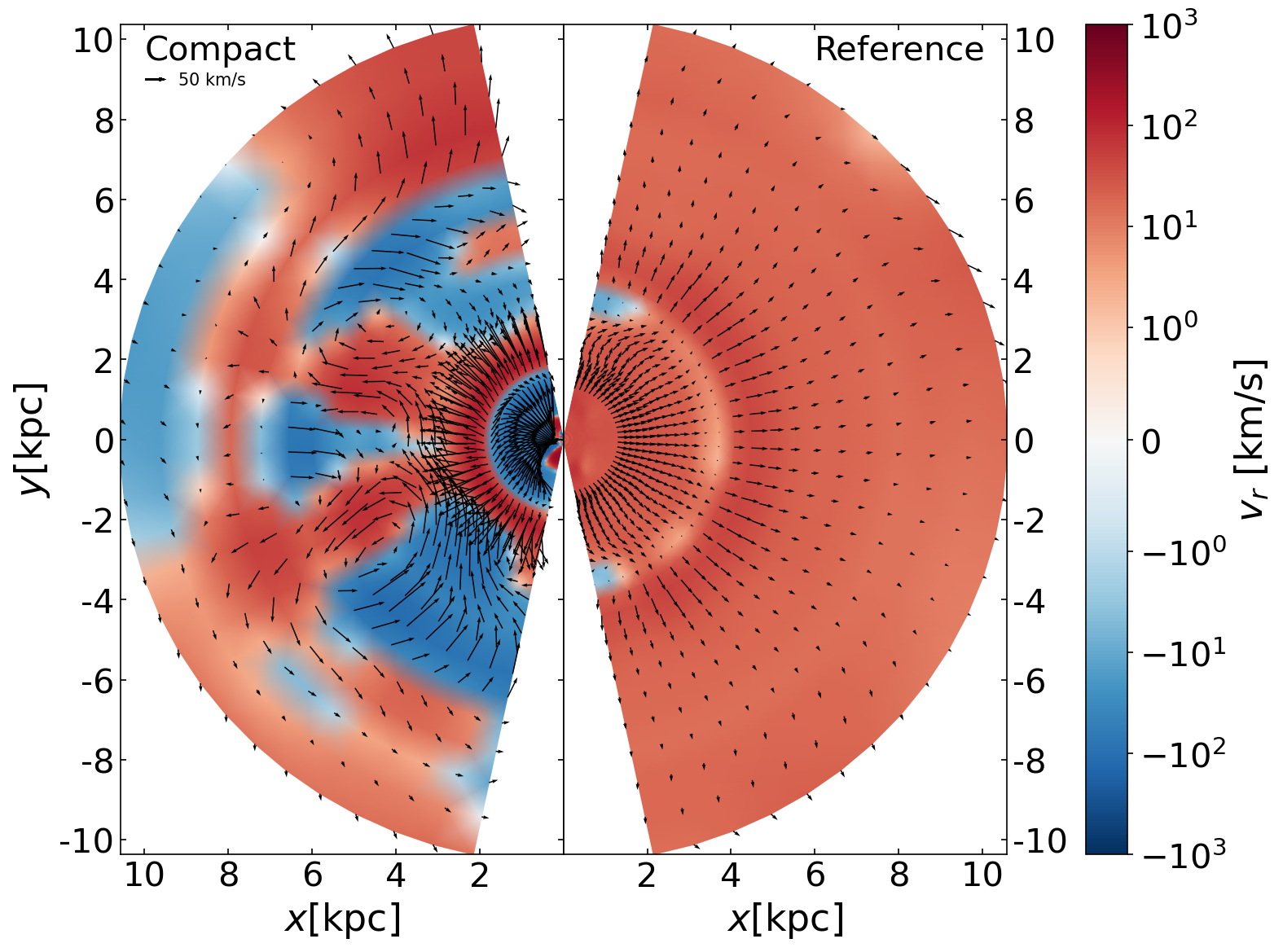}}
\caption{
Panel (a): The radial velocity of gas for the compact (left) and reference (right) galaxies averaged over the entire 10 Gyr period. 
Panel (b): The radial velocity in the two galaxies at 9.75 Gyr within the central 10 kpc. Arrows represent the velocity vector of the gas. This snapshot was randomly selected. }
\label{fig:vmap}
\end{figure*}

\begin{figure*}  
\centering
\includegraphics[width=15cm]{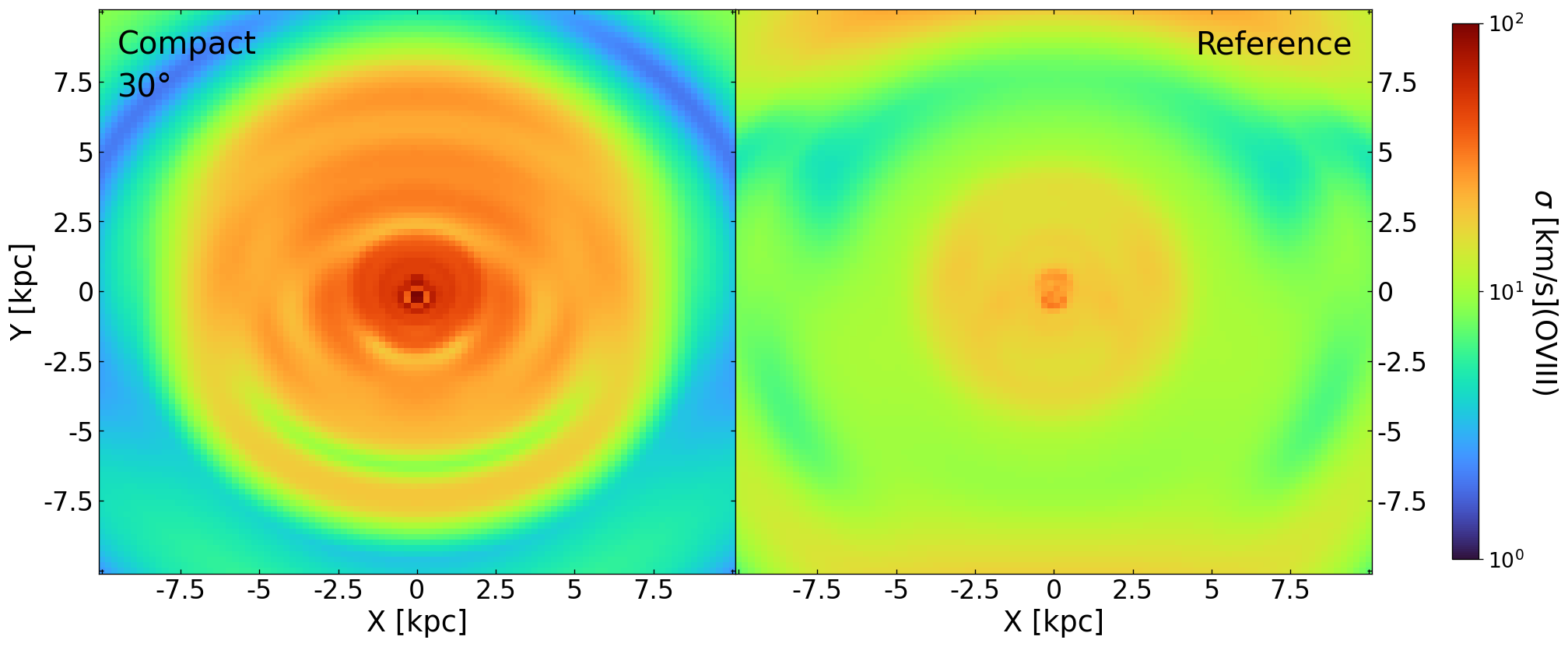}
\caption{ The velocity dispersion ($\sigma$) of O VIII-emitting gas in the compact ({\it left}) and reference ({\it right}) galaxies at 9.75 Gyr in the simulation, corresponding to the same snapshot of Figure~\ref{fig:snap}. $\sigma$ is weighted by O VIII Ly$\alpha$ emissivity. The rings are likely caused by the limit of 2D simulations. In 3D simulations, these symmetric rings will likely break apart into clumpy features.
}
\label{fig:sigma_map}
\end{figure*}

Figure~\ref{fig:avr} shows the radial velocity distribution for the compact and reference galaxies averaged over the entire 10 Gyr period. In both cases, the flow forms a fountain-like structure, with the outflow mainly in the polar region and the inflow along the equatorial plane. Such a structure is caused by the bipolar structure of wind launched from the AGN. In the compact galaxy, the fountain flow extends to a few kpc, while in the reference galaxy, the fountain pattern is confined to the central $\sim0.01$ kpc. This difference is because the black hole accretion rate in the compact galaxy is about 10 times larger than the reference galaxy, as we discuss in more detail in section \ref{sec:star}, thus the wind in the compact galaxy is much stronger than that in the reference galaxy.  On average, both the inflow and the outflow in the compact galaxy have higher speeds than those in the reference galaxy. In both galaxies, even though galactic-scale bi-polar outflows can form at the peaks of AGN outbursts, most of the time the galaxy hosts a subsonic galactic wind with no preferred directions.
Figure~\ref{fig:snap} shows the distribution of the velocity field for the compact and reference galaxies at 9.75 Gyr. This snapshot was randomly selected and displayed with linear spacing in radius. 
The fountain-like structure remains in the compact galaxy and the flow velocities are larger in the compact galaxy.

The predicted difference between the compact and reference galaxies can be tested by observing heavy ion emission lines in X-ray spectra. Figure \ref{fig:sigma_map} shows the predicted velocity dispersion of O VIII-emitting gas in the compact and reference galaxies corresponding to Figure \ref{fig:snap}. The line of sight is adopted to be $30$ degrees. To produce the mock observations, we first produced the 3D data based on our 2D simulation data by assuming axial symmetry. We then compute O VIII emissivity based on ATOMDB\footnote{\url{www.atomdb.org}}.
This figure shows that the compact galaxy has a much higher velocity dispersion than the reference galaxy.
This prediction can be tested using the upcoming high-resolution X-ray instruments such as {\it Athena} X-IFU \citep{2018SPIE10699E..1GB} or {\it HUBS} \citep{2020JLTP..199..502C}. 
When warm gas exists, the predictions could also be observed by the Integral Field Spectrographs (IFS) on some optical telescopes. Note that the rings shown in the figure are caused by torus-like structure within which the typical gas velocity is significantly lower than the gas outside of the torus. Such a 3D torus-like structure corresponds to some vortexes in the 2D simulation data, so we expect they should be replaced by more tangled structure in the more realistic 3D simulation.

\subsection{Star Formation and SMBH Accretion Rate}\label{sec:star}

\begin{figure}  
\centering
\includegraphics[width=8cm]{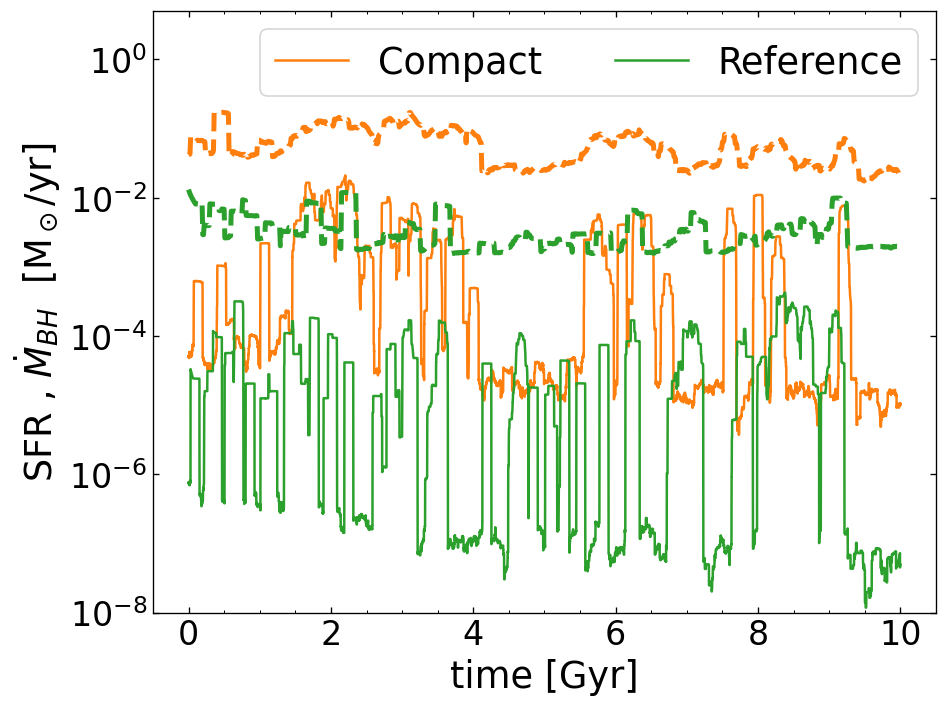}
\caption{The evolution of SMBH accretion rate (dashed lines) and SFR (solid lines) of the simulated compact (orange) and reference (green) galaxies. For clarity, all rates are smoothed with a moving window of 200 Myr.}
\label{fig:sfr}
\end{figure}

Figure \ref{fig:sfr} shows the evolution of the star formation rate (SFR) and black hole accretion rate $\dot{M}_{\rm BH}$ in compact and reference galaxies. The SFR in the compact galaxy (solid orange line) is roughly two orders of magnitude higher than the reference galaxy due to its higher gas density. But the SFR in the compact galaxy is still very low, with a median $\sim 10^{-4} \mathrm{M_\odot}/yr$ and the highest value below 1 $\rm M_{\odot}/yr$.  The corresponding specific SFR are $10^{-15}\,\mathrm{yr^{-1}}$ and $10^{-11}\,\mathrm{yr^{-1}}$, respectively, both of which being lower than $0.3/t_{\rm hubble}$. In other words, there is some very low-level star formation activity, and the simulated galaxy stays quiescent by any observational standards \citep[e.g.,][]{2008ApJ...688..770F}. The average spatial distribution of the newly formed stars roughly follows that of the existing stellar population, similar to the reference galaxy and what has been found in previous simulations of massive elliptical galaxies \citep{Yuan2018,Wang2019}. 

The SMBH accretion rate in the compact galaxy is higher than the reference galaxy by a factor of $\sim$10. The AGN in the compact galaxy is in the cold mode about 0.08\% of the time, while for the reference galaxy, it is 0.02\%. Over the 10 Gyr evolution period, the SMBH mass grows by $\sim 50\%$ in the compact galaxy, compared with $\sim 3\%$ in the reference galaxy. Some observational works \citep[e.g.,][]{2015ApJ...808..183W, 2016ApJ...817....2W} have suggested that SMBHs in local compact galaxies tend to be overly massive given the host's luminosity and bulge mass \citep[although see][]{2021ApJ...919...77C}. The SMBH mass growth in our simulations is overall insignificant. If a larger observational sample in the future confirms that compact galaxies do tend to host overly more massive SMBHS, this trend is likely set by the formation history of the galaxies in the distant past, rather than the late-time SMBH growth at lower redshifts.

\section{Conclusions and Final Remarks}\label{sec:conclusions}

We have performed high-resolution two-dimensional numerical simulations of SMBH feeding and feedback in an idealized compact galaxy in the framework of {\it MACER} and compared the results with a simulated typical low-z elliptical galaxy. We study the SMBH-driven outflow and the accretion flow from galaxy scales down to below the Bondi radius. The compact galaxy shows larger inflow and outflow rates compared with the reference galaxy, which is mainly because the gas density in the compact galaxy is higher due to its more compact stellar distribution. The high inflow rate in the compact galaxy results in strong SMBH feeding and feedback. Consequently, the compact galaxy forms a large galaxy-scale inflow-outflow fountain, while the reference galaxy shows a relatively smooth and gentle outflow. Due to the stronger SMBH feeding, we find that compact galaxies tend to have stronger winds. We observe this in the form of faster velocities and higher velocity dispersions. This prediction can be potentially tested with IFS observations.

Multi-phase gas can develop within a few kpc about 10\% of the time, following the typical precipitation criteria. However, the accretion flow past the inner boundary within the Bondi radius is always single phase, as the cool/warm gas is mixed back to the hot phase before reaching the SMBH. 

Due to the higher gas density in the compact galaxy, its SFR is roughly two orders of magnitude larger than the reference galaxy. But during the entire 10 Gyr simulation duration, the compact galaxy still stays quiescent with very low-level star formation activities. Due to its stronger inflow rate, the SMBH grows by $\sim 50\%$ in the compact galaxy, significantly larger than that in the reference galaxy, which is $\sim 3\%$.  

\section*{Acknowledgments}
We would like to thank Lei Hao, Suoqing Ji, Miao Li, Zhiyuan Li, and Mark Voit for helpful discussions. We thank the referee for his/her constructive suggestions which significantly improved the paper.
YD, FY and FS are supported in part by the NSF of China (grants 12133008, 12192220, and 12192223),
YL is supported in part by the NSF (grants AST-2107735 and AST-2219686). The calculations have made use of the High Performance Computing Resource in the Core Facility for Advanced Research Computing at Shanghai Astronomical Observatory.
\section*{Data Availability}
The data underlying this article will be shared on reasonable request to the corresponding author.

\bibliographystyle{mnras}
\bibliography{RedNuggets.bib}

\begin{thebibliography}{}
\makeatletter
\relax
\def\mn@urlcharsother{\let\do\@makeother \do\$\do\&\do\#\do\^\do\_\do\%\do\~}
\def\mn@doi{\begingroup\mn@urlcharsother \@ifnextchar [ {\mn@doi@}
  {\mn@doi@[]}}
\def\mn@doi@[#1]#2{\def\@tempa{#1}\ifx\@tempa\@empty \href
  {http://dx.doi.org/#2} {doi:#2}\else \href {http://dx.doi.org/#2} {#1}\fi
  \endgroup}
\def\mn@eprint#1#2{\mn@eprint@#1:#2::\@nil}
\def\mn@eprint@arXiv#1{\href {http://arxiv.org/abs/#1} {{\tt arXiv:#1}}}
\def\mn@eprint@dblp#1{\href {http://dblp.uni-trier.de/rec/bibtex/#1.xml}
  {dblp:#1}}
\def\mn@eprint@#1:#2:#3:#4\@nil{\def\@tempa {#1}\def\@tempb {#2}\def\@tempc
  {#3}\ifx \@tempc \@empty \let \@tempc \@tempb \let \@tempb \@tempa \fi \ifx
  \@tempb \@empty \def\@tempb {arXiv}\fi \@ifundefined
  {mn@eprint@\@tempb}{\@tempb:\@tempc}{\expandafter \expandafter \csname
  mn@eprint@\@tempb\endcsname \expandafter{\@tempc}}}

\bibitem[\protect\citeauthoryear{{Barret} et~al.,}{{Barret}
  et~al.}{2018}]{2018SPIE10699E..1GB}
{Barret} D.,  et~al., 2018, in {den Herder} J.-W.~A.,  {Nikzad} S.,
  {Nakazawa} K.,  eds,  Society of Photo-Optical Instrumentation Engineers
  (SPIE) Conference Series Vol. 10699, Space Telescopes and Instrumentation
  2018: Ultraviolet to Gamma Ray. p. 106991G (\mn@eprint {arXiv} {1807.06092}),
  \mn@doi{10.1117/12.2312409}

\bibitem[\protect\citeauthoryear{{Bu}, {Yuan}, {Gan}  \& {Yang}}{{Bu}
  et~al.}{2016a}]{2016ApJ...818...83B}
{Bu} D.-F.,  {Yuan} F.,  {Gan} Z.-M.,   {Yang} X.-H.,  2016a, \mn@doi [\apj]
  {10.3847/0004-637X/818/1/83}, \href
  {https://ui.adsabs.harvard.edu/abs/2016ApJ...818...83B} {818, 83}

\bibitem[\protect\citeauthoryear{{Bu}, {Yuan}, {Gan}  \& {Yang}}{{Bu}
  et~al.}{2016b}]{2016ApJ...823...90B}
{Bu} D.-F.,  {Yuan} F.,  {Gan} Z.-M.,   {Yang} X.-H.,  2016b, \mn@doi [\apj]
  {10.3847/0004-637X/823/2/90}, \href
  {https://ui.adsabs.harvard.edu/abs/2016ApJ...823...90B} {823, 90}

\bibitem[\protect\citeauthoryear{{Buote} \& {Barth}}{{Buote} \&
  {Barth}}{2018}]{2018ApJ...854..143B}
{Buote} D.~A.,  {Barth} A.~J.,  2018, \mn@doi [\apj]
  {10.3847/1538-4357/aaa971}, \href
  {https://ui.adsabs.harvard.edu/abs/2018ApJ...854..143B} {854, 143}

\bibitem[\protect\citeauthoryear{{Cheung} et~al.,}{{Cheung}
  et~al.}{2016}]{Cheung2016}
{Cheung} E.,  et~al., 2016, \mn@doi [\nat] {10.1038/nature18006}, \href
  {https://ui.adsabs.harvard.edu/abs/2016Natur.533..504C} {533, 504}

\bibitem[\protect\citeauthoryear{{Ciotti} \& {Ostriker}}{{Ciotti} \&
  {Ostriker}}{2001}]{2001ApJ...551..131C}
{Ciotti} L.,  {Ostriker} J.~P.,  2001, \mn@doi [\apj] {10.1086/320053}, \href
  {https://ui.adsabs.harvard.edu/abs/2001ApJ...551..131C} {551, 131}

\bibitem[\protect\citeauthoryear{{Ciotti}, {Ostriker}  \& {Proga}}{{Ciotti}
  et~al.}{2010}]{2010ApJ...717..708C}
{Ciotti} L.,  {Ostriker} J.~P.,   {Proga} D.,  2010, \mn@doi [\apj]
  {10.1088/0004-637X/717/2/708}, \href
  {https://ui.adsabs.harvard.edu/abs/2010ApJ...717..708C} {717, 708}

\bibitem[\protect\citeauthoryear{{Cohn} et~al.,}{{Cohn}
  et~al.}{2021}]{2021ApJ...919...77C}
{Cohn} J.~H.,  et~al., 2021, \mn@doi [\apj] {10.3847/1538-4357/ac0f78}, \href
  {https://ui.adsabs.harvard.edu/abs/2021ApJ...919...77C} {919, 77}

\bibitem[\protect\citeauthoryear{{Cui} et~al.,}{{Cui}
  et~al.}{2020}]{2020JLTP..199..502C}
{Cui} W.,  et~al., 2020, \mn@doi [Journal of Low Temperature Physics]
  {10.1007/s10909-019-02279-3}, \href
  {https://ui.adsabs.harvard.edu/abs/2020JLTP..199..502C} {199, 502}

\bibitem[\protect\citeauthoryear{{Damjanov} et~al.,}{{Damjanov}
  et~al.}{2009}]{2009ApJ...695..101D}
{Damjanov} I.,  et~al., 2009, \mn@doi [\apj] {10.1088/0004-637X/695/1/101},
  \href {https://ui.adsabs.harvard.edu/abs/2009ApJ...695..101D} {695, 101}

\bibitem[\protect\citeauthoryear{Dehnen}{Dehnen}{1993}]{Dehnen1993}
Dehnen W.,  1993, \mn@doi [\mnras] {10.1093/mnras/265.1.250}, 265, 250

\bibitem[\protect\citeauthoryear{{Fabian}}{{Fabian}}{2012}]{2012ARA&A..50..455F}
{Fabian} A.~C.,  2012, \mn@doi [\araa] {10.1146/annurev-astro-081811-125521},
  \href {https://ui.adsabs.harvard.edu/abs/2012ARA&A..50..455F} {50, 455}

\bibitem[\protect\citeauthoryear{{Fang}, {Faber}, {Koo}  \& {Dekel}}{{Fang}
  et~al.}{2013}]{2013ApJ...776...63F}
{Fang} J.~J.,  {Faber} S.~M.,  {Koo} D.~C.,   {Dekel} A.,  2013, \mn@doi [\apj]
  {10.1088/0004-637X/776/1/63}, \href
  {https://ui.adsabs.harvard.edu/abs/2013ApJ...776...63F} {776, 63}

\bibitem[\protect\citeauthoryear{{Franx}, {van Dokkum}, {F{\"o}rster
  Schreiber}, {Wuyts}, {Labb{\'e}}  \& {Toft}}{{Franx}
  et~al.}{2008}]{2008ApJ...688..770F}
{Franx} M.,  {van Dokkum} P.~G.,  {F{\"o}rster Schreiber} N.~M.,  {Wuyts} S.,
  {Labb{\'e}} I.,   {Toft} S.,  2008, \mn@doi [\apj] {10.1086/592431}, \href
  {https://ui.adsabs.harvard.edu/abs/2008ApJ...688..770F} {688, 770}

\bibitem[\protect\citeauthoryear{Gaspari, Ruszkowski  \& Oh}{Gaspari
  et~al.}{2013}]{Gaspari2013}
Gaspari M.,  Ruszkowski M.,   Oh S.~P.,  2013, \mn@doi [\mnras]
  {10.1093/mnras/stt692}, 432, 3401

\bibitem[\protect\citeauthoryear{Gofford, Reeves, McLaughlin, Braito, Turner,
  Tombesi  \& Cappi}{Gofford et~al.}{2015}]{Gofford2015}
Gofford J.,  Reeves J.~N.,  McLaughlin D.~E.,  Braito V.,  Turner T.~J.,
  Tombesi F.,   Cappi M.,  2015, \mn@doi [\mnras] {10.1093/mnras/stv1207}, 451,
  4169

\bibitem[\protect\citeauthoryear{{Guo}, {Stone}, {Kim}  \& {Quataert}}{{Guo}
  et~al.}{2022}]{Guo2022}
{Guo} M.,  {Stone} J.~M.,  {Kim} C.-G.,   {Quataert} E.,  2022, arXiv e-prints,
  \href {https://ui.adsabs.harvard.edu/abs/2022arXiv221105131G} {p.
  arXiv:2211.05131}

\bibitem[\protect\citeauthoryear{Ji, Oh  \& McCourt}{Ji et~al.}{2018}]{Ji2018}
Ji S.,  Oh S.~P.,   McCourt M.,  2018, \mn@doi [Mon. Not. Roy. Astron. Soc.]
  {10.1093/mnras/sty293}, 476, 852

\bibitem[\protect\citeauthoryear{{Kriek}, {van Dokkum}, {Labb{\'e}}, {Franx},
  {Illingworth}, {Marchesini}  \& {Quadri}}{{Kriek}
  et~al.}{2009}]{2009ApJ...700..221K}
{Kriek} M.,  {van Dokkum} P.~G.,  {Labb{\'e}} I.,  {Franx} M.,  {Illingworth}
  G.~D.,  {Marchesini} D.,   {Quadri} R.~F.,  2009, \mn@doi [\apj]
  {10.1088/0004-637X/700/1/221}, \href
  {https://ui.adsabs.harvard.edu/abs/2009ApJ...700..221K} {700, 221}

\bibitem[\protect\citeauthoryear{{Li} \& {Bryan}}{{Li} \&
  {Bryan}}{2014}]{2014ApJ...789..153L}
{Li} Y.,  {Bryan} G.~L.,  2014, \mn@doi [\apj] {10.1088/0004-637X/789/2/153},
  \href {https://ui.adsabs.harvard.edu/abs/2014ApJ...789..153L} {789, 153}

\bibitem[\protect\citeauthoryear{{Ma}, {Roberts}, {Li}  \& {Wang}}{{Ma}
  et~al.}{2019}]{2019MNRAS.483.5614M}
{Ma} R.-Y.,  {Roberts} S.~R.,  {Li} Y.-P.,   {Wang} Q.~D.,  2019, \mn@doi
  [\mnras] {10.1093/mnras/sty3039}, \href
  {https://ui.adsabs.harvard.edu/abs/2019MNRAS.483.5614M} {483, 5614}

\bibitem[\protect\citeauthoryear{{Maraston}}{{Maraston}}{2005}]{2005MNRAS.362..799M}
{Maraston} C.,  2005, \mn@doi [\mnras] {10.1111/j.1365-2966.2005.09270.x},
  \href {https://ui.adsabs.harvard.edu/abs/2005MNRAS.362..799M} {362, 799}

\bibitem[\protect\citeauthoryear{{McNamara} \& {Nulsen}}{{McNamara} \&
  {Nulsen}}{2007}]{2007ARA&A..45..117M}
{McNamara} B.~R.,  {Nulsen} P.~E.~J.,  2007, \mn@doi [\araa]
  {10.1146/annurev.astro.45.051806.110625}, \href
  {https://ui.adsabs.harvard.edu/abs/2007ARA&A..45..117M} {45, 117}

\bibitem[\protect\citeauthoryear{{Naab}, {Johansson}  \& {Ostriker}}{{Naab}
  et~al.}{2009}]{2009ApJ...699L.178N}
{Naab} T.,  {Johansson} P.~H.,   {Ostriker} J.~P.,  2009, \mn@doi [\apjl]
  {10.1088/0004-637X/699/2/L178}, \href
  {https://ui.adsabs.harvard.edu/abs/2009ApJ...699L.178N} {699, L178}

\bibitem[\protect\citeauthoryear{{Narayan}, {S{\"A} dowski}, {Penna}  \&
  {Kulkarni}}{{Narayan} et~al.}{2012}]{2012MNRAS.426.3241N}
{Narayan} R.,  {S{\"A} dowski} A.,  {Penna} R.~F.,   {Kulkarni} A.~K.,  2012,
  \mn@doi [\mnras] {10.1111/j.1365-2966.2012.22002.x}, \href
  {https://ui.adsabs.harvard.edu/abs/2012MNRAS.426.3241N} {426, 3241}

\bibitem[\protect\citeauthoryear{Navarro, Frenk  \& White}{Navarro
  et~al.}{1996}]{Navarro1996}
Navarro J.~F.,  Frenk C.~S.,   White S. D.~M.,  1996, \mn@doi [\apj]
  {10.1086/177173}, 462, 563

\bibitem[\protect\citeauthoryear{{Negri} \& {Volonteri}}{{Negri} \&
  {Volonteri}}{2017}]{2017MNRAS.467.3475N}
{Negri} A.,  {Volonteri} M.,  2017, \mn@doi [\mnras] {10.1093/mnras/stx362},
  \href {https://ui.adsabs.harvard.edu/abs/2017MNRAS.467.3475N} {467, 3475}

\bibitem[\protect\citeauthoryear{{Novak}, {Ostriker}  \& {Ciotti}}{{Novak}
  et~al.}{2011}]{2011ApJ...737...26N}
{Novak} G.~S.,  {Ostriker} J.~P.,   {Ciotti} L.,  2011, \mn@doi [\apj]
  {10.1088/0004-637X/737/1/26}, \href
  {https://ui.adsabs.harvard.edu/abs/2011ApJ...737...26N} {737, 26}

\bibitem[\protect\citeauthoryear{Oser, Ostriker, Naab, Johansson  \&
  Burkert}{Oser et~al.}{2010}]{Oser2010}
Oser L.,  Ostriker J.~P.,  Naab T.,  Johansson P.~H.,   Burkert A.,  2010,
  \mn@doi [\apj] {10.1088/0004-637x/725/2/2312}, 725, 2312

\bibitem[\protect\citeauthoryear{{Park}, {Hada}, {Kino}, {Nakamura}, {Ro}  \&
  {Trippe}}{{Park} et~al.}{2019}]{Park2019}
{Park} J.,  {Hada} K.,  {Kino} M.,  {Nakamura} M.,  {Ro} H.,   {Trippe} S.,
  2019, \mn@doi [\apj] {10.3847/1538-4357/aaf9a9}, \href
  {https://ui.adsabs.harvard.edu/abs/2019ApJ...871..257P} {871, 257}

\bibitem[\protect\citeauthoryear{{Pillepich} et~al.,}{{Pillepich}
  et~al.}{2018}]{2018MNRAS.473.4077P}
{Pillepich} A.,  et~al., 2018, \mn@doi [\mnras] {10.1093/mnras/stx2656}, \href
  {https://ui.adsabs.harvard.edu/abs/2018MNRAS.473.4077P} {473, 4077}

\bibitem[\protect\citeauthoryear{{Prasad}, {Voit}, {O'Shea}  \&
  {Glines}}{{Prasad} et~al.}{2020}]{2020ApJ...905...50P}
{Prasad} D.,  {Voit} G.~M.,  {O'Shea} B.~W.,   {Glines} F.,  2020, \mn@doi
  [\apj] {10.3847/1538-4357/abc33c}, \href
  {https://ui.adsabs.harvard.edu/abs/2020ApJ...905...50P} {905, 50}

\bibitem[\protect\citeauthoryear{Ressler, Quataert  \& Stone}{Ressler
  et~al.}{2018}]{Ressler2018}
Ressler S.~M.,  Quataert E.,   Stone J.~M.,  2018, \mn@doi [\mnras]
  {10.1093/mnras/sty1146}, 478, 3544

\bibitem[\protect\citeauthoryear{{Schaye} et~al.,}{{Schaye}
  et~al.}{2015}]{2015MNRAS.446..521S}
{Schaye} J.,  et~al., 2015, \mn@doi [\mnras] {10.1093/mnras/stu2058}, \href
  {https://ui.adsabs.harvard.edu/abs/2015MNRAS.446..521S} {446, 521}

\bibitem[\protect\citeauthoryear{Sharma, McCourt, Quataert  \& Parrish}{Sharma
  et~al.}{2012}]{Sharma2012}
Sharma P.,  McCourt M.,  Quataert E.,   Parrish I.~J.,  2012, \mn@doi [\mnras]
  {10.1111/j.1365-2966.2011.20246.x}, 420, 3174

\bibitem[\protect\citeauthoryear{{Shi}, {Li}, {Yuan}  \& {Zhu}}{{Shi}
  et~al.}{2021}]{Shi2021}
{Shi} F.,  {Li} Z.,  {Yuan} F.,   {Zhu} B.,  2021, \mn@doi [Nature Astronomy]
  {10.1038/s41550-021-01394-0}, \href
  {https://ui.adsabs.harvard.edu/abs/2021NatAs...5..928S} {5, 928}

\bibitem[\protect\citeauthoryear{{Shi}, {Zhu}, {Li}  \& {Yuan}}{{Shi}
  et~al.}{2022}]{2022ApJ...926..209S}
{Shi} F.,  {Zhu} B.,  {Li} Z.,   {Yuan} F.,  2022, \mn@doi [\apj]
  {10.3847/1538-4357/ac4789}, \href
  {https://ui.adsabs.harvard.edu/abs/2022ApJ...926..209S} {926, 209}

\bibitem[\protect\citeauthoryear{{Stone} \& {Norman}}{{Stone} \&
  {Norman}}{1992}]{1992ApJS...80..753S}
{Stone} J.~M.,  {Norman} M.~L.,  1992, \mn@doi [\apjs] {10.1086/191680}, \href
  {https://ui.adsabs.harvard.edu/abs/1992ApJS...80..753S} {80, 753}

\bibitem[\protect\citeauthoryear{Szomoru, Franx  \& van Dokkum}{Szomoru
  et~al.}{2012}]{Szomoru2012}
Szomoru D.,  Franx M.,   van Dokkum P.~G.,  2012, \mn@doi [\apj]
  {10.1088/0004-637x/749/2/121}, 749, 121

\bibitem[\protect\citeauthoryear{{Tacchella} et~al.,}{{Tacchella}
  et~al.}{2015}]{2015Sci...348..314T}
{Tacchella} S.,  et~al., 2015, \mn@doi [Science] {10.1126/science.1261094},
  \href {https://ui.adsabs.harvard.edu/abs/2015Sci...348..314T} {348, 314}

\bibitem[\protect\citeauthoryear{{Trujillo}, {Ferr{\'e}-Mateu}, {Balcells},
  {Vazdekis}  \& {S{\'a}nchez-Bl{\'a}zquez}}{{Trujillo}
  et~al.}{2014}]{2014ApJ...780L..20T}
{Trujillo} I.,  {Ferr{\'e}-Mateu} A.,  {Balcells} M.,  {Vazdekis} A.,
  {S{\'a}nchez-Bl{\'a}zquez} P.,  2014, \mn@doi [\apjl]
  {10.1088/2041-8205/780/2/L20}, \href
  {https://ui.adsabs.harvard.edu/abs/2014ApJ...780L..20T} {780, L20}

\bibitem[\protect\citeauthoryear{{Voit}, {Kay}  \& {Bryan}}{{Voit}
  et~al.}{2005}]{Voit2005}
{Voit} G.~M.,  {Kay} S.~T.,   {Bryan} G.~L.,  2005, \mn@doi [\mnras]
  {10.1111/j.1365-2966.2005.09621.x}, \href
  {https://ui.adsabs.harvard.edu/abs/2005MNRAS.364..909V} {364, 909}

\bibitem[\protect\citeauthoryear{{Voit}, {Donahue}, {O'Shea}, {Bryan}, {Sun}
  \& {Werner}}{{Voit} et~al.}{2015}]{Voit2015}
{Voit} G.~M.,  {Donahue} M.,  {O'Shea} B.~W.,  {Bryan} G.~L.,  {Sun} M.,
  {Werner} N.,  2015, \mn@doi [\apjl] {10.1088/2041-8205/803/2/L21}, \href
  {https://ui.adsabs.harvard.edu/abs/2015ApJ...803L..21V} {803, L21}

\bibitem[\protect\citeauthoryear{Voit, Meece, Li, O'Shea, Bryan  \&
  Donahue}{Voit et~al.}{2017}]{Voit2017}
Voit G.~M.,  Meece G.,  Li Y.,  O'Shea B.~W.,  Bryan G.~L.,   Donahue M.,
  2017, \mn@doi [\apj] {10.3847/1538-4357/aa7d04}, 845, 80

\bibitem[\protect\citeauthoryear{{Walsh}, {van den Bosch}, {Gebhardt},
  {Yildirim}, {G{\"u}ltekin}, {Husemann}  \& {Richstone}}{{Walsh}
  et~al.}{2015}]{2015ApJ...808..183W}
{Walsh} J.~L.,  {van den Bosch} R. C.~E.,  {Gebhardt} K.,  {Yildirim} A.,
  {G{\"u}ltekin} K.,  {Husemann} B.,   {Richstone} D.~O.,  2015, \mn@doi [\apj]
  {10.1088/0004-637X/808/2/183}, \href
  {https://ui.adsabs.harvard.edu/abs/2015ApJ...808..183W} {808, 183}

\bibitem[\protect\citeauthoryear{{Walsh}, {van den Bosch}, {Gebhardt},
  {Y{\i}ld{\i}r{\i}m}, {Richstone}, {G{\"u}ltekin}  \& {Husemann}}{{Walsh}
  et~al.}{2016}]{2016ApJ...817....2W}
{Walsh} J.~L.,  {van den Bosch} R. C.~E.,  {Gebhardt} K.,  {Y{\i}ld{\i}r{\i}m}
  A.,  {Richstone} D.~O.,  {G{\"u}ltekin} K.,   {Husemann} B.,  2016, \mn@doi
  [\apj] {10.3847/0004-637X/817/1/2}, \href
  {https://ui.adsabs.harvard.edu/abs/2016ApJ...817....2W} {817, 2}

\bibitem[\protect\citeauthoryear{{Wang} et~al.,}{{Wang}
  et~al.}{2013}]{wang2013}
{Wang} Q.~D.,  et~al., 2013, \mn@doi [Science] {10.1126/science.1240755}, \href
  {https://ui.adsabs.harvard.edu/abs/2013Sci...341..981W} {341, 981}

\bibitem[\protect\citeauthoryear{{Wang}, {Li}  \& {Ruszkowski}}{{Wang}
  et~al.}{2019}]{Wang2019}
{Wang} C.,  {Li} Y.,   {Ruszkowski} M.,  2019, \mn@doi [\mnras]
  {10.1093/mnras/sty2906}, \href
  {https://ui.adsabs.harvard.edu/abs/2019MNRAS.482.3576W} {482, 3576}

\bibitem[\protect\citeauthoryear{{Weinberger} et~al.,}{{Weinberger}
  et~al.}{2018}]{2018MNRAS.479.4056W}
{Weinberger} R.,  et~al., 2018, \mn@doi [\mnras] {10.1093/mnras/sty1733}, \href
  {https://ui.adsabs.harvard.edu/abs/2018MNRAS.479.4056W} {479, 4056}

\bibitem[\protect\citeauthoryear{{Werner}, {Lakhchaura}, {Canning}, {Gaspari}
  \& {Simionescu}}{{Werner} et~al.}{2018}]{2018MNRAS.477.3886W}
{Werner} N.,  {Lakhchaura} K.,  {Canning} R.~E.~A.,  {Gaspari} M.,
  {Simionescu} A.,  2018, \mn@doi [\mnras] {10.1093/mnras/sty862}, \href
  {https://ui.adsabs.harvard.edu/abs/2018MNRAS.477.3886W} {477, 3886}

\bibitem[\protect\citeauthoryear{Xie \& Yuan}{Xie \& Yuan}{2012}]{Xie2012}
Xie F.-G.,  Yuan F.,  2012, \mn@doi [\apj] {10.1111/j.1365-2966.2012.22030.x},
  427, 1580

\bibitem[\protect\citeauthoryear{{Yang}, {Yuan}, {Yuan}  \& {White}}{{Yang}
  et~al.}{2021}]{Yang2021}
{Yang} H.,  {Yuan} F.,  {Yuan} Y.-F.,   {White} C.~J.,  2021, \mn@doi [\apj]
  {10.3847/1538-4357/abfe63}, \href
  {https://ui.adsabs.harvard.edu/abs/2021ApJ...914..131Y} {914, 131}

\bibitem[\protect\citeauthoryear{{Y{\i}ld{\i}r{\i}m}, {van den Bosch}, {van de
  Ven}, {Mart{\'\i}n-Navarro}, {Walsh}, {Husemann}, {G{\"u}ltekin}  \&
  {Gebhardt}}{{Y{\i}ld{\i}r{\i}m} et~al.}{2017}]{2017MNRAS.468.4216Y}
{Y{\i}ld{\i}r{\i}m} A.,  {van den Bosch} R. C.~E.,  {van de Ven} G.,
  {Mart{\'\i}n-Navarro} I.,  {Walsh} J.~L.,  {Husemann} B.,  {G{\"u}ltekin} K.,
    {Gebhardt} K.,  2017, \mn@doi [\mnras] {10.1093/mnras/stx732}, \href
  {https://ui.adsabs.harvard.edu/abs/2017MNRAS.468.4216Y} {468, 4216}

\bibitem[\protect\citeauthoryear{Yuan \& Narayan}{Yuan \&
  Narayan}{2014}]{Yuan2014}
Yuan F.,  Narayan R.,  2014, \mn@doi [\araa]
  {10.1146/annurev-astro-082812-141003}, 52, 529

\bibitem[\protect\citeauthoryear{{Yuan}, {Wu}  \& {Bu}}{{Yuan}
  et~al.}{2012a}]{2012ApJ...761..129Y}
{Yuan} F.,  {Wu} M.,   {Bu} D.,  2012a, \mn@doi [\apj]
  {10.1088/0004-637X/761/2/129}, \href
  {https://ui.adsabs.harvard.edu/abs/2012ApJ...761..129Y} {761, 129}

\bibitem[\protect\citeauthoryear{{Yuan}, {Bu}  \& {Wu}}{{Yuan}
  et~al.}{2012b}]{2012ApJ...761..130Y}
{Yuan} F.,  {Bu} D.,   {Wu} M.,  2012b, \mn@doi [\apj]
  {10.1088/0004-637X/761/2/130}, \href
  {https://ui.adsabs.harvard.edu/abs/2012ApJ...761..130Y} {761, 130}

\bibitem[\protect\citeauthoryear{Yuan, Gan, Narayan, Sadowski, Bu  \& Bai}{Yuan
  et~al.}{2015}]{Yuan2015}
Yuan F.,  Gan Z.,  Narayan R.,  Sadowski A.,  Bu D.,   Bai X.-N.,  2015,
  \mn@doi [\apj] {10.1088/0004-637x/804/2/101}, 804, 101

\bibitem[\protect\citeauthoryear{Yuan, Yoon, Li, Gan, Ho  \& Guo}{Yuan
  et~al.}{2018}]{Yuan2018}
Yuan F.,  Yoon D.,  Li Y.-P.,  Gan Z.-M.,  Ho L.~C.,   Guo F.,  2018, \mn@doi
  [\apj] {10.3847/1538-4357/aab8f8}, 857, 121

\bibitem[\protect\citeauthoryear{{Zolotov} et~al.,}{{Zolotov}
  et~al.}{2015}]{2015MNRAS.450.2327Z}
{Zolotov} A.,  et~al., 2015, \mn@doi [\mnras] {10.1093/mnras/stv740}, \href
  {https://ui.adsabs.harvard.edu/abs/2015MNRAS.450.2327Z} {450, 2327}

\bibitem[\protect\citeauthoryear{{van Dokkum} et~al.,}{{van Dokkum}
  et~al.}{2010}]{2010ApJ...709.1018V}
{van Dokkum} P.~G.,  et~al., 2010, \mn@doi [\apj]
  {10.1088/0004-637X/709/2/1018}, \href
  {https://ui.adsabs.harvard.edu/abs/2010ApJ...709.1018V} {709, 1018}

\bibitem[\protect\citeauthoryear{van Dokkum et~al.,}{van Dokkum
  et~al.}{2015}]{Dokkum2015}
van Dokkum P.~G.,  et~al., 2015, \mn@doi [\apj] {10.1088/0004-637x/813/1/23},
  813, 23

\makeatother
\end{thebibliography}

\bsp	
\label{lastpage}
\end{document}